\documentclass[iop]{emulateapj}
\usepackage{epsfig}
\usepackage{subfigure}
\usepackage{mathtools}

\def\Kepler{\textit{Kepler}}
\def\Rearth{R_\oplus}

\def\Pmax{150}
\def\npps{2.00}
\def\enpps{0.45}

\begin{document}

\title{The Radius Distribution of Planets Around Cool Stars}
\author{Timothy D. Morton\altaffilmark{1,2}, Jonathan Swift\altaffilmark{2}}
\email{tdm@astro.princeton.edu}
\altaffiltext{1}{Department of Astrophysical Sciences, 4 Ivy Lane, Peyton Hall, 
Princeton University, Princeton, NJ 08544}
\altaffiltext{2}{Department of Astrophysics,
  California Institute of Technology, MC 249-17, Pasadena, CA 91125}

\begin{abstract}

We calculate an empirical, non-parametric estimate of the shape of the period-marginalized radius distribution of planets with periods less than 150 days using the small yet well-characterized sample of cool ($T_{\rm eff} <4000 $K) dwarf stars in the \Kepler\ catalog.  In particular, we present and validate a new procedure, based on weighted kernel density estimation, to reconstruct the shape of the planet radius function down to radii smaller than the completeness limit of the survey at the longest periods.  Under the assumption that the period distribution of planets does not change dramatically with planet radius, we show that the occurrence of planets around these stars continues to increase to below 1 $\Rearth$, and that there is no strong evidence for a turnover in the planet radius function. In fact, we demonstrate using many iterations of simulated data that a spurious turnover may be inferred from data even when the true distribution continues to rise toward smaller radii.  Finally, the sharp rise in the radius distribution below $\sim$3 $\Rearth$ implies that a large number of planets await discovery around cool dwarfs as the sensitivities of ground-based transit surveys increase.

\end{abstract}

\section{Introduction}

The discovery of the first exoplanets \citep{wol92,mayor1995,marcy1996} has sparked
tremendous growth in research and interest in the formation and
evolution of planetary systems beyond the Solar System. Most of the first few dozen extrasolar planets found had masses greater than Saturn and semimajor axes less than 0.5 AU. But not unlike
many areas of astronomy, the first discoveries are not representative
samples; rather, close-in giant planets are relatively rare
\citep{wright2012,howard2010} in comparison to the new populations of exoplanets now
being revealed by the \Kepler\ Mission \citep{borucki2011,batalha2012,burke2013}. The
most common kinds of planets within \Kepler's discovery space
of $R_p \gtrsim 0.5 \Rearth$, and $P \lesssim 100$\,d appear to
be somewhat larger than Earth but smaller than Neptune, $1 < R_p
<4 \Rearth$ \citep{howard2012,fressin2013,dressing2013}.

Much of our understanding of planet formation is anchored in decades
of research into our own Solar System.  But now the burgeoning exoplanet 
population provides us with a new context revealing important insights
into planet formation throughout the Galaxy. For example, the large amount of planetary mass seen close to host stars 
is evidence that protoplanetary disks may have much higher surface
densities than previously thought \citep{hansen2012,chiang2012} or that the observed planets migrated
from regions further from their host star where more mass was readily
available for assembly \citep{swift2013,raymond2014}.

The architectures of Kepler planetary systems also offer a wealth of information regarding their formation and evolution.  The masses of a subset of planets in Kepler's multi systems have been measured from the effects of mutual gravitational interactions (e.g. Cochran et al. 2011, Lithwick et al. 2012, Nesvorny et al. 2013) providing insight into the composition and atmospheric evolution of these planets \citep{wu2013,owen2013,lopez2012}.  While more recently, individual planet masses are being measured with precise radial velocity observations \citep{marcy2014},  advancing our understanding of the compositional makeup of the Kepler planet sample \citep[e.g.,][]{weiss2014}.  The low inferred mutual inclination of multi-transit systems \citep[$\sim 1^\circ$--$3^\circ$;][]{fabrycky2012b,fang2012}
together with the relative number of single versus multi-transit systems provides constraints
on the number of planets in a given system within \Kepler's discovery
window, else it may be the first indication of a separate,
high-inclination population of single transit systems \citep{hansen2013,fang2012}.

In this article we focus on yet another important clue regarding the
formation of the compact systems revealed by \Kepler: the
distribution of planetary radii.  Constructing the underlying distribution of planet radii from the results of a transit survey is a subtle task.  Early in the history of transit surveys, \citet{gaudi2005} and \citet{gaudietal2005} identified many of the pitfalls inherent to this endeavor---mainly the strong period and radius biases.  While the characteristics of the \Kepler\ survey make it less susceptible to these issues, these biases still persist and need to be accounted for.  

The radius distribution of planets as derived from the \Kepler\ survey has been the subject of a number of studies in the literature.  The initial estimates of the planet
radius distribution by Howard et al. (2012) showed a dramatic increase
in the number of planets at ever smaller size. Citing incompleteness, however, 
they did not follow this trend in their analysis to planet radii smaller
than 2 $\Rearth$. In an independent study by \citet{youdin2011}, a
parametric estimation of the planetary distribution function revealed a deficit of large planets in short period orbits that would support a formation scenario involving core accretion followed by inward migration; \citet{dong2012} report a similar finding.

More recent estimates of the planet radius distribution show a
preferential size scale in the \Kepler\ sample indicated by
a flattening and possible turnover in the log-binned histogram of detected planet candidates
somewhere around 2\ $\Rearth$ \citep{fressin2013,dressing2013,petigura2013,petigura2013b}. If true, this would be an important clue toward understanding the key
mechanisms that shape the observed population of compact planetary 
systems that pervade the Galaxy.  However, these analyses are constrained by the limitations of coarse histograms; no analysis to date has yet characterized the shape of the exoplanet radius distribution in enough detail to allow meaningful comparison to planet formation and evolution theories.   Additionally, all of the above except for \citet{dressing2013} rely on the \Kepler\ Input Catalog for the stellar properties of the target star population, the known unreliability of which has been shown to bias the results of statistical analyses \citep{gaidos2013}.  

The goal of the present study is distinct from previous work in several ways.  First, we focus exclusively on the smallest stars in the \Kepler\ target sample as a well characterized subsample, due to both the photometric re-calibration of stellar properties presented by \citet{dressing2013} and because the host stars of many of the \Kepler\ Objects of Interest (KOIs) in this sample have been investigated spectroscopically \citep[Muirhead et al., in prep;][]{muirhead2012a}.  Second, we aim to reconstruct as faithfully as possible the detailed shape of the radius function, avoiding both the limitations of histogram binning and the assumption that the distribution follows a power law---a non-parametric, non-histogram approach to this problem has not yet been attempted.  And finally, in order to investigate any potential flattening or turnover of the distribution, we extend planet occurrence analysis to radii smaller than has been attempted before, introducing a new technique that allows proper marginalization over period even for radii for which the \Kepler\ survey is beginning to be incomplete.

In \S\ref{sec:formalism} we walk through the steps required to extract this non-parametric empirical estimate of the true period-marginalized planet radius function given a population of detections from a well-characterized transit survey.  In \S\ref{sec:calculation} we apply these methods to the Cool KOIs to derive the radius distribution for small planets around small stars down to below 1 $\Rearth$. We summarize our results in \S\ref{sec:results} and explore the  assumptions that go into this calculation in \S\ref{sec:discussion}.  Concluding remarks follow in \S\ref{sec:conclusion}.

\section{Formalism}
\label{sec:formalism}

We define the planet radius distribution function $\phi_r^{P_{\rm max}}(r)$ such that
\begin{eqnarray}
\label{eq:definition}
& \int_{r_{\rm min}}^{r_{\rm max}}  \phi^{P_{\rm max}}_r(r) dr = {\rm NPPS,}  ~P < P_{\rm max};   
\end{eqnarray}
that is, a density function with an overall normalization giving the average number of planets per star (NPPS) for planets with period less than $P_{\rm max}$ days, for planet radii $r$ between $r_{\rm min}$ and $r_{\rm max}$.  The problem of calculating planet occurrence rates from \Kepler\ has been quite an industry over the last few years \citep{youdin2011,howard2012,dong2012,swift2013,fressin2013,petigura2013,dressing2013}.  However, there has been little quantitative discussion of deriving the detailed shape of the radius function beyond drawing histograms.  In the following subsections, we review and refine the general principles of an occurrence calculation and then describe how to follow these principles to construct a non-parametric empirical radius function that obeys the above desired properties.

\subsection{Occurrence Calculations}
\label{sec:occurrence}
In a perfectly idealized survey that is both 100\% reliable and 100\% complete, the occurrence rate of planets (in a survey, or in a specified bin) is simply 
\begin{equation}
\label{eq:npps}
{\rm NPPS} = \frac{N_p}{N_\star},
\end{equation}  
where $N_p$ is the number of detected planets and $N_\star$ is the number of stars surveyed.  In practice, however, this must be corrected for both incompleteness and unreliability as follows:

\begin{equation}
\label{eq:nppssum}
{\rm NPPS} = \frac{1}{N_{\star}} \sum_{i=1}^{N_p} w_i.
\end{equation} 
Here the sum is over all detections and $w_i$ is a weighting factor applied individually to account for the various necessary corrections.  Generally, these weights can be thought of as

\begin{equation}
\label{eq:weightsimple}
w_i = \frac{(1-{\rm FPP}_i)}{\eta_i},
\end{equation}
where ${\rm FPP}_i$ is the probability that signal $i$ is a false positive and $\eta_i$ is an individualized efficiency factor for the detection of planet $i$.  In this work, we calculate the ${\rm FPP}_i$ according to the procedure in \citet{morton2012}; see \S\ref{sec:fpp} for details.

We thus focus discussion here on the detection efficiency $\eta_i$, which is defined by the following thought experiment:  \emph{If a very large number of planets identical to planet $i$ were distributed randomly around all the stars in the survey, only a fraction $\eta_i$ could have been detected.}  Conceptually, this can be factored \citep[following][]{youdin2011}:
\begin{equation}
\label{eq:etafactored}
\eta_i = \eta_{{\rm tr},i} \cdot \eta_{{\rm disc},i},
\end{equation} 
where $\eta_{\rm tr}$ is the geometric transit probability, and $\eta_{\rm disc}$ is the ``discovery efficiency'': the fraction of planets in this thought experiment \emph{with transiting orbital geometries} that could have been detected by the survey. Previous \Kepler\ occurrence rate calculations \citep{howard2012,swift2013,dressing2013}, have defined the discovery efficiency as
\begin{equation}
\label{eq:Nstar}
\eta_{{\rm disc},i} = \frac{N_{\star,i}}{N_{\star}}
\end{equation}
where $N_{\star,i}$ is the number of target stars around which planet $i$ could have been detected.  This number is determined by counting the stars around which a transit of planet $i$ (at period $P_i$) would cause a photometric signal with signal-to-noise ratio (SNR) above some detection threshold (10 for \citet{howard2012}, 7.1 for \citet{dressing2013} and \citet{swift2013}).  

Though not spelled out explicitly, this way of calculating $\eta_{{\rm disc},i}$ essentially boils down to two general ingredients:  a hypothetical ``ensemble-of-alternative-scenarios'' SNR probability distribution $\phi_{{\rm SNR},i}$ for each planet in the survey, and the discovery efficiency as a function of SNR $\eta_{\rm SNR}$ for the transit detection pipeline.  Given these two ingredients, \citet{howard2012}, \citet{dressing2013}, and \citet{swift2013} all implicitly calculate the following:
\begin{equation}
\label{eq:etadisc}
\eta_{{\rm disc},i} = \int_0^\infty  \eta_{\rm SNR}(s) \cdot \phi_{{\rm SNR},i}(s) ds,
\end{equation} 
where $\phi_{{\rm SNR},i}$ is a normalized probability distribution of SNR ($s$) for that planet determined by varying the stellar radius and noise properties according to each star in the survey, and $\eta_{\rm SNR}$ is assumed to be a step function at some detection threshold.

It is important to note, however, that a step function  does not accurately characterize the true detection efficiency of the \Kepler\ pipeline as a function of SNR.  \citet{fressin2013} found that the observed SNR distribution of the \citet{batalha2012} catalog implied that the true behavior was more like an ``SNR ramp,'' where the discovery efficiency was zero at $SNR=6$ and unity at $SNR=16$.  Internal tests of the \Kepler\ pipeline indicate that the true shape of the $\eta_{\rm SNR}$ function is similar to this (P.~Tennenbaum, priv.~comm., based on poster at the 2nd \Kepler\ Science Conference).  An accurate estimate of this function is crucial to obtaining correct results in an occurrence rate calculation; we present the functional form we use in \S\ref{sec:calculation}.

\begin{figure}[t!]
  \centering
   \includegraphics[width=3.5in]{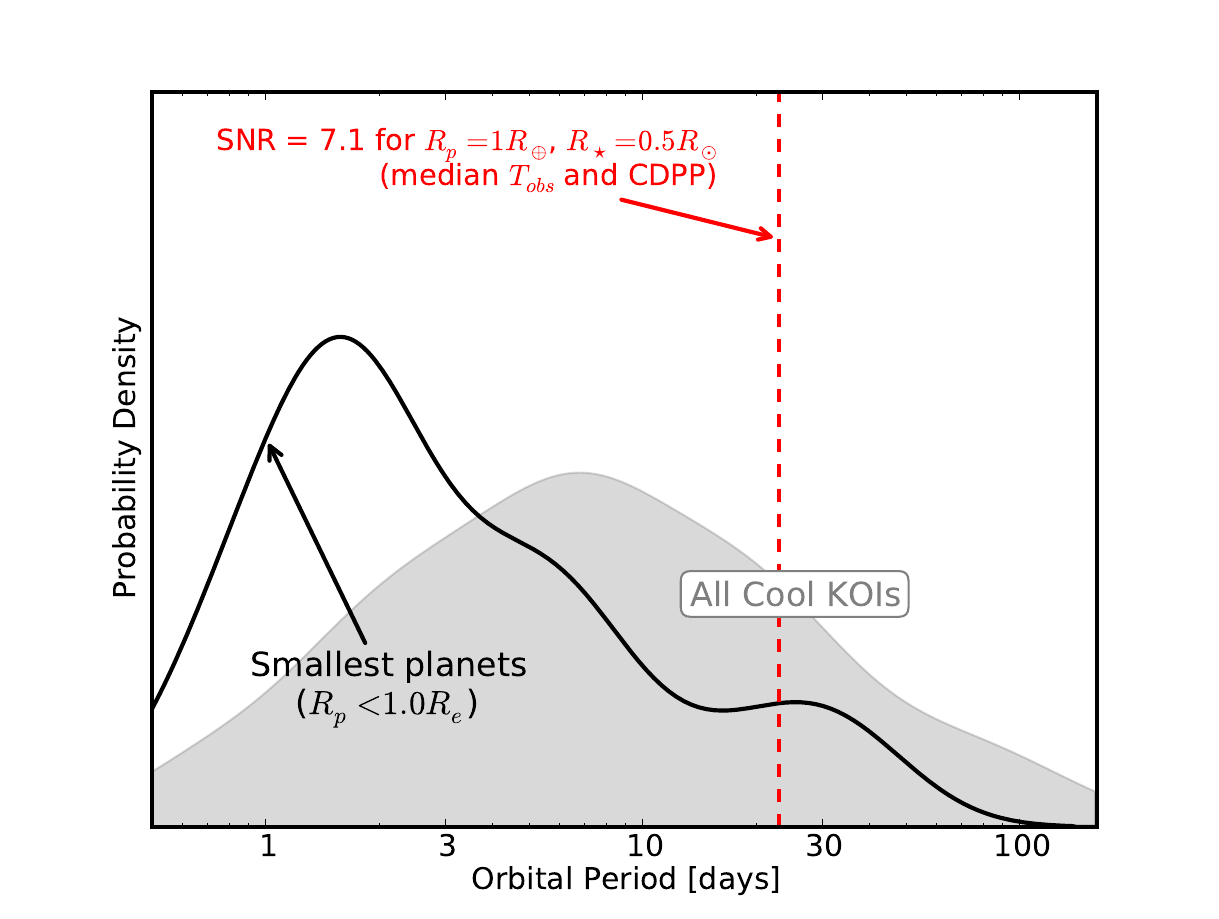} 
   \caption{Evidence supporting the hypothesis that small planets are incomplete in the Cool KOI sample.  The solid black line is the observed period distribution of planets smaller than 1 $\Rearth$; the grey shaded area is the observed period distribution of all the Cool KOIs.  (Neither distribution is corrected for transit probability.)  The vertical dashed red line indicates the period at which a 1 $\Rearth$ planet around a 0.5 $R_\odot$ star (typical of the Cool KOI sample) would have SNR of 7.1, the nominal detection threshold for KOI identification.  The relative lack of observed small planets at periods longer than 10 days is thus very plausibly due to incompleteness.}
    \label{fig:pincomplete}
\end{figure}

We also note that the traditional method of computing $\phi_{{\rm SNR},i}$---simulating a planet of radius $r_i$ transiting all of the survey stars at fixed period $P_i$---is designed to estimate the \textit{joint period-radius distribution} of planets.  This joint distribution is then summed up over period to find the period-marginalized radius distribution---this is how planet occurrence as a function of radius is determined in \citet{howard2012}, \citet{dressing2013}, \citet{petigura2013}, and \citet{petigura2013b}.  However, as Figure \ref{fig:pincomplete} demonstrates, the smallest planets in the Cool KOI survey are likely only complete out to periods of a few tens of days, whereas many larger planets are detected on larger orbits.  Thus, determining the period-marginalized radius function of planets using this traditional approach requires either restricting the analysis to radii above which the survey is substantially complete out to $P=P_{\rm max}$, or only using a small fraction of the survey detections (out to periods of only a few tens of days) in order to probe below $\sim$1 $\Rearth$.  

We introduce here an alternative method of computing $\phi_{{\rm SNR},i}$ that allows for better reconstruction of the planet radius function at small radii without having to restrict analysis to a fraction of the available data.  Instead of simulating the ensemble of alternative transit scenarios for planet $i$ all at the single fixed period $P_i$, we suggest that these alternative scenarios could be assigned a distribution of periods according to a reasonable estimate of the true planet period distribution.  This effectively amounts to a strategy of ``pre-marginalization'' that requires an assumption of the period distribution of planets but allows for every planet detection to be treated on an equal basis.

Making this adjustment to how $\eta_{{\rm disc},i}$ is determined---by distributing hypothetical planets around all stars \textit{at all periods} to calculate $\phi_{{\rm SNR},i}$ ---also necessitates rethinking how transit probability is corrected for.  That is,  construction of $\phi_{{\rm SNR},i}$ must acknowledge that transit probability (as well as SNR) depends on planet period and host star radius.  In other words, any process of building up a hypothetical SNR distribution from many instances of simulated transits must ensure that each instance is properly weighted by its actual likelihood of detection.  The most straightforward way of ensuring this is to simulate an isotropic cloud of planets around each target star and include only the transiting configurations (e.g., impact parameter $b <= 1$) in the SNR distribution.  The transit probability factor $\eta_{{\rm tr},i}$ then becomes the fraction of all these simulated planets that transit---this will typically \textit{not} be the same as the individual geometric transit probability of planet $i$, since planet $i$ has a single period $P_i$ while the simulated population has a distribution of periods.  In other words, calculating the radius function pre-marginalized over period in this way demands that only a single survey- and period-averaged transit probability be used when calculating the completeness correction for each planet---a function of only the assumed period distribution and the distribution of stellar radii of the target sample.

\subsection{Estimating the Radius Distribution Function}
\label{sec:radfunction}

In all the \Kepler\ planet occurrence calculations to date, the shape of the radius function has been explored only very coarsely, by calculating the occurrence rate in several different radius bins and either fitting a power law or qualitatively commenting on the shape.  \citet{howard2012} found a good fit to an $R^{-2}$ power law down to 2 $\Rearth$, and declined to comment for smaller planets.  On the other hand, \citet{fressin2013} and \citet{petigura2013} note that the occurrence rate of planets increases towards smaller radius but then appears to flatten out below about 2.8 $\Rearth$.  \citet{dressing2013} claim that the occurrence rate begins to decrease for planets  smaller than 1-1.4 $\Rearth$, and \citet{petigura2013b} find that the planet occurrence rate decreases for planets smaller than 2 $\Rearth$. 

Investigating the shape of the radius distribution in more detail requires a non-parametric approach, and also should avoid binning.  Here we introduce the concept of using a weighted kernel density estimator in order to accomplish this.

\subsubsection{Weighted Kernel Density Estimation}
\label{sec:wkde}

A standard kernel density estimator (KDE) attempts to estimate the true underlying probability distribution of a sample of data points using a function of the following form:
\begin{equation}
\label{eq:kde}
\hat\phi(x) = \frac{1}{N} \sum_{i=1}^N k(x-x_i;\sigma_i),
\end{equation}
where $N$ is the number of data points and $k(x)$ is a zero-mean, normalized kernel function of arbitrary shape (commonly a Gaussian), with some bandwidth $\sigma_i$, that most generally can be different for each data point.  This creates a smooth distribution out of a discrete data set, with the degree of smoothness controlled by the width parameter.  The choice of width has tradeoffs in both directions: if the kernels are too narrow the estimator will be bumpy, but if they are too wide they can wash out real structure in the distribution.  Often the width is selected to be the same for all points based on the number of data points, or sometimes a variable-width kernel is used, e.g.~the distance to the $n$th nearest neighbor.  The $1/N$ normalization factor assures that the integral of this density estimator over the whole parameter space is unity. 

In order to use the KDE concept to properly reconstruct the radius function of planets detected in a transit survey, each data point has to be weighted appropriately, leading to a weighted KDE, or wKDE:
\begin{equation}
\label{eq:wkde}
\hat\phi^{P_{\rm max}}_r(r) = \frac{1}{N_\star} \sum_{i=1}^{N_p}  w_i \cdot k(r-r_i;\sigma_i),
\end{equation}
where $w_i = 1/\eta_i$ are the appropriately calculated individual weight factors that renormalize the kernels to correct for missing planets, as discussed in \S\ref{sec:occurrence}.  The weights ensure that the shape of the radius function responds appropriately to the individual corrections, and the $N_\star$ overall normalization ensures that the integral over all radii will return the NPPS, as desired in Equation \ref{eq:definition}.  We note that because this function does not normalize to unity, it is not strictly a density function, but a ``rate function,'' representing number of planets per unit radius rather than probability density.  

\citet{wang2007} explore in detail several techniques for selecting the optimal kernel bandwidths, presenting in particular two methods of the form 
\begin{equation}
\label{eq:wangh}
\sigma_{\rm opt} = 0.9 C n^{-1/5}
\end{equation}
where $C$ is a constant based on the sample mean, sample variance, or interquartile range of the data, and $n$ is the number of data points.  However, we show in \S\ref{sec:calculation} that for our particular case, using the actual observational uncertainties in planet radius to control the width of the individual smoothing kernels $\sigma_i$ results in a smooth distribution, so we do not apply this technique.  We do note, however, that if the planet radii were known more precisely, then a method such as this would be necessary in order to quantify an optimum smoothing width to avoid high-frequency features in the estimator.

\subsubsection{De-biasing the wKDE and calculating variance}
\label{sec:biasvariance}

\begin{figure}[t!]
   \centering
   \includegraphics[width=3.5in]{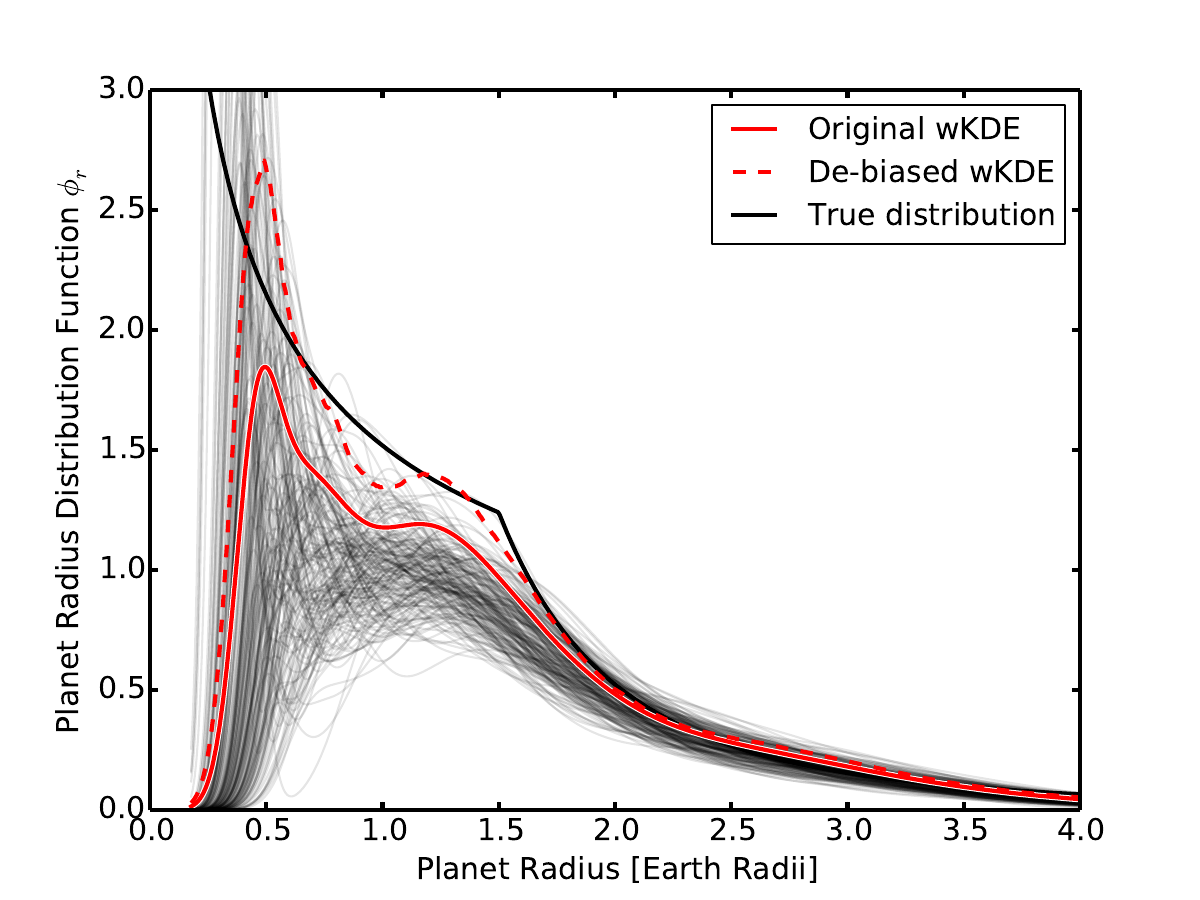} 
   \caption{An illustration of the de-biasing procedure for the wKDE estimator discussed in this work.  Mock transit survey data is simulated, with the size of the planets following the distribution illustrated by the thick black line.  From this data, a wKDE of the radius distribution is derived using the methods described in \S\ref{sec:wkde} (solid red line).  Then, 200 new transit-survey datasets are generated with the planet radii drawn from the estimated red-line distribution, and corresponding new wKDEs are generated according to the same procedure (thin black lines).  The thin black lines tend to fall systematically below the solid red line; this is an illustration of the bias of this estimator.  This difference is added back to the red line to produce the de-biased estimate of the distribution (dashed red line), which matches the true distribution more closely.  The spread in the bootstrap estimators around their median illustrates the estimator's variance.  This technique is known as the ``smoothed bootstrap'' \citep{narsky2013}.}
      \label{fig:debiasing}
\end{figure}

The wKDE $\hat\phi_r$ described in Equation \ref{eq:wkde} is an estimator of the true underlying radius distribution $\phi_r$, and like any estimator, it has both bias and variance associated with it.  These quantities must be determined in order to best understand what the data can tell us about the true distribution.  \citet{narsky2013} explain how to calculate bias and variance for a standard KDE using a resampling technique called the ``smoothed bootstrap''; we adopt this method for our purpose and describe it here.  

While a traditional bootstrap technique involves resampling the observed dataset with replacement to create new datasets, a ``smoothed bootstrap'' involves generating new datasets \textit{according to the estimated distribution}.  The density estimator is recalculated for each of these simulated datasets using the same procedure that generated the original estimator.  The median offset of these resampled estimators relative to the the original estimator (which can be directly observed) will then reflect the bias of the original estimator relative to the true underlying distribution, and the original estimator can thus be corrected accordingly.  Similarly, the variance of the estimator can be determined by the scatter of these bootstrapped estimators.  

Applying this method to a wKDE in the context of a transit survey is less straightforward than simple resampling according to the derived $\hat\phi_r$, but we borrow the same principle.  Each smoothed bootstrap resampling dataset is generated by simulating a whole new survey: drawing a set of planets according to the estimated distribution, assigning them isotropically distributed orbits around host stars drawn randomly from the target sample, calculating the SNR of each of the resulting transits, and then determining which of these planets would be detectable (using the appropriate $\eta_{\rm SNR}$ function).  Each of these datasets is then converted into a new wKDE, and the bias may thus be assessed.

Figure \ref{fig:debiasing} illustrates this procedure applied in a toy scenario.  The true underlying planet radius distribution---a broken power law---is illustrated with the thick solid black line.  The initial wKDE estimate of the radius distribution derived from one realization of ``detected'' planets is shown as a solid red line.  Thin black lines illustrate 200 wKDEs resulting from resampled datasets generated according to the originally calculated wKDE distribution (red line).  It is clear that for below 1.5 $\Rearth$ or so, the bootstrap resamplings underestimate the red line; this mimics the way the red line is biased with respect to the true distribution.  So to correct for this, the difference between the solid red line and the median of the bootstrapped distributions is added to the the initial wKDE estimate at each value of radius to obtain the de-biased wKDE (dashed red line), which matches the true distribution more closely.  The error region around this de-biased estimate is then taken to be the same as the spread in the bootstrapped wKDEs about their median.  

In addition to illustrating how the bias and variance of these wKDEs may be calculated, this example also touches on two other important points.  First, this procedure demonstrates that naive summing of weights to calculate NPPS (Equation \ref{eq:npps}) is in fact most generally \textit{a biased estimator} for the true NPPS---a non-intuitive but significant result.  And second, the toy-model true radius distribution used here is a power law that continues to rise all the way down to 0.3 $\Rearth$.  However, because of the detection sensitivity of the survey, only very rarely are any planets smaller than 0.5 $\Rearth$ detected.  Thus, there is generically a turnover in the estimated planet radius distribution---the location of which depends on which planets happen to be observed.  In fact, the estimated distribution sometimes even appears to turn over around $\sim$1 $\Rearth$---despite the fact that the true distribution continues to rise.  Recognizing this is crucial to properly interpreting the radius function derived from \Kepler\ data; we return to discussion of this point in \S\ref{sec:discussion}.

\section{Calculating the Cool KOI Radius Function}
\label{sec:calculation}

One of the biggest concerns to date about interpreting \Kepler\ data is uncertainty about stellar parameters.  This applies both because the properties of the transit host stars are unknown (derived planet radius depends directly on the radius of the host star) and because the properties of the stars in the survey parent sample are unknown (i.e.~is \Kepler\ actually surveying dwarf stars or is the parent sample significantly contaminated by giants or subgiants? \citep{mann2012}).

Focusing on \Kepler\ candidates around relatively low-mass stars alleviates these concerns.  Many of these stars have spectroscopically measured stellar properties \citep{muirhead2012b,mann2012}, and in addition, the properties of the parent sample of target stars has been carefully characterized photometrically by \citet{dressing2013}.  Such an investigation thus is narrower than attempting to use the whole \Kepler\ sample, but the assurance of a good understanding of the stellar parameters of both the host stars and the general survey sample more than compensates for this loss of generality.  In addition, focusing on these ``Cool KOIs'' enables detailed study of the radius distribution of Earth-sized and smaller planets.  

To construct the planet radius function, we thus select the 130 \Kepler\ Objects of Interest (KOIs) with periods $<$$\Pmax$d  identified in the  Q1-Q12 KOI catalog posted at the NASA Exoplanet Archive that are hosted by stars with $T_{\rm eff} < 4000$K as characterized by \citet{dressing2013}.  To this stellar sample we add KOI-961/Kepler-42, which was left out of the \citet{dressing2013} sample because its broad-band colors are consistent with classification
as either a giant or a dwarf, even though it has been spectroscopically confirmed to be a $\sim$0.15 $M_\odot$ dwarf \citep{muirhead2012a}.   (We do note, however, that KOI-961.02 is not in this Q1-Q12 KOI catalog because the minimum period in the Q1-Q12 \Kepler\ pipeline search was 0.5 days.)  
For stellar parameters we use the results presented in \citet{dressing2013}, except for those KOI host stars that have been spectroscopically characterized according to the observations and procedures described in  \citet{muirhead2012a}, for which we use the spectroscopic parameters (Muirhead et al., in prep).

In the following subsections, we describe the steps necessary to calculate $\hat\phi^{\Pmax}_{r}$, the estimate of the radius function for planets on orbits $<${\Pmax}d, from this KOI sample.  As described in \S\ref{sec:formalism}, the crucial step toward properly estimating the radius function is calculating the weight factor $w_i = 1/\eta_i$ for each detection, which includes a transit probability factor and a completeness factor $\eta_{{\rm disc},i}$ (Equation \ref{eq:etadisc}).  Key to calculating $\eta_{{\rm disc},i}$ is determining the SNR distribution of a hypothetical population of clones of planet $i$ around all the target stars, or $\phi_{{\rm SNR},i}$.  To accomplish this, we use the ``pre-marginalization'' strategy as explained in \S\ref{sec:occurrence}, which in turn requires an assumption of the intrinsic period distribution of planets $\phi_P$.

\begin{figure}[t!]
   \centering
   \includegraphics[width=3.5in]{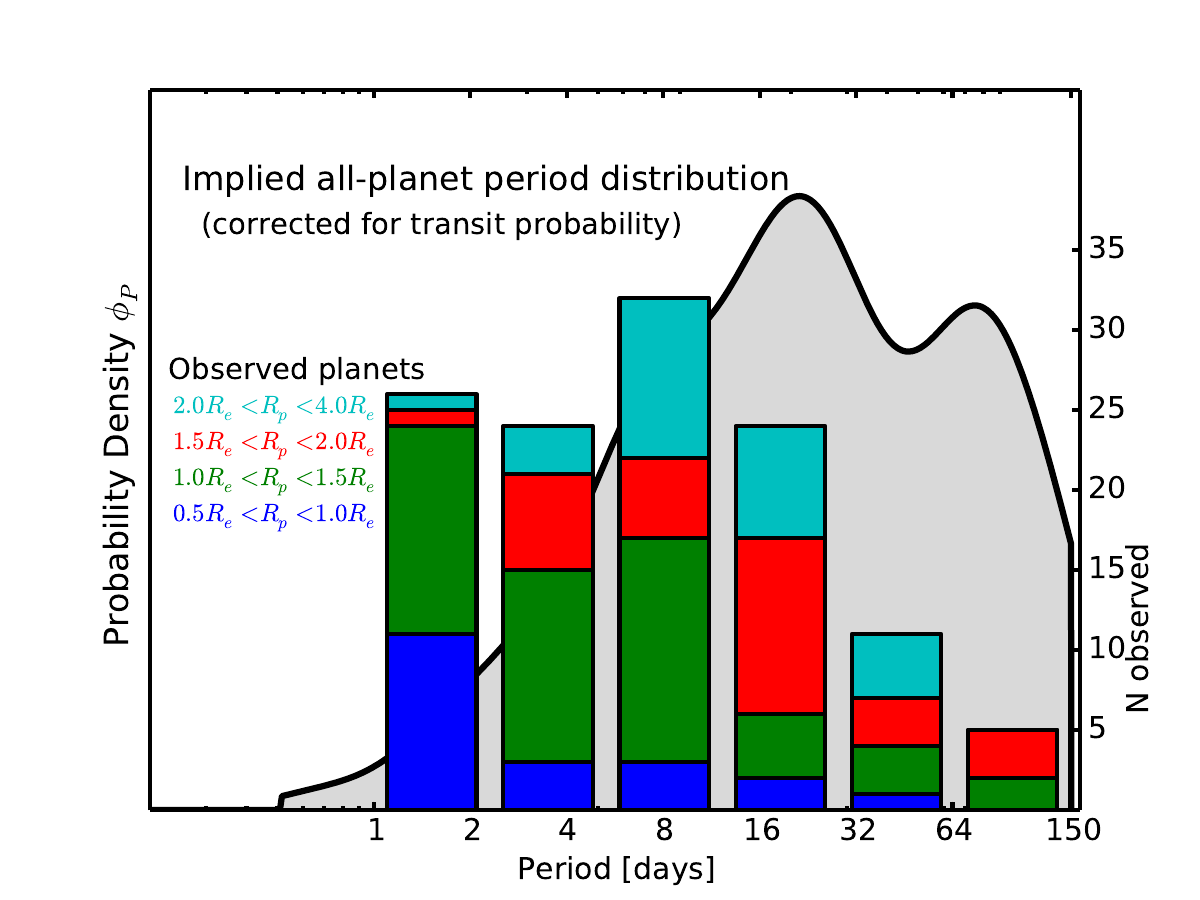} 
   \caption{The period distribution of planets around \Kepler's M dwarfs.  The grey shaded region is the implied period distribution of all planets combined, correcting for the effects of transit probability.  The bar charts show the observed numbers of planets of different sizes in each period bin.  Note the declining fraction of small planets as a function of period---this is most likely an effect of declining detection efficiency for smaller planets on longer-period orbits, and this must be properly accounted for when constructing the period-marginalized planet radius function down to small radii.  The radius function calculation in this paper assumes that all planets are distributed according to the shaded distribution, regardless of planet radius.  See \S\ref{sec:perdistassumption} for a discussion of this assumption.}
   \label{fig:pdist}
\end{figure}

\subsection{Period distribution}
\label{sec:perdist}

In order to estimate the shape of the true period distribution of planets of all sizes, we make the simplifying assumption that the period distribution of planets is independent of their radii (see \S\ref{sec:perdistassumption} for discussion regarding this assumption).  We thus construct the distribution of $\log P$ from all the planet candidates in the sample, using a wKDE  as described in \S\ref{sec:radfunction}.   For the weights we use only the inverse transit probabilities, and enforce that the whole distribution is normalized to unity, creating the probability density function for $\log P$.  For the widths we use $\sigma = 0.15$ (in $\log P$), to create a smooth distribution.  This is the period distribution function $\phi_P$ that we use in the following subsection, shown as the grey shaded region in Figure \ref{fig:pdist}.

\subsection{SNR distribution}
\label{sec:snrdist}

The SNR of a transit signal is usually defined as follows:
\begin{equation}
\label{eq:snr}
{\rm SNR} = \frac{\delta}{\sigma} \sqrt{N_{\rm tr} \cdot N_{\rm pts}},
\end{equation}
where $\delta$ is the transit depth, $\sigma$ is the one-point photometric uncertainty, $N_{\rm tr}$ is the number of transits observed, and $N_{\rm pts}$ is the number of photometric points per transit.

In order to construct $\phi_{{\rm SNR},i}$, the distribution of SNRs for every conceivable alternative scenario in which planet of radius $r_i$ might have existed in this survey (around every other target star at any other potential period, under our pre-marginalization strategy; see \S\ref{sec:occurrence}), we use a Monte Carlo simulation.  For each target star, 50,000 planets of radius $r_i$ are generated with isotropic orbital inclinations and orbital periods according to the period distribution of \S\ref{sec:perdist}.  We then calculate the transit signal SNR for each planet in this simulation that has a transiting impact parameter, with $\delta$ being the depth of a \citet{ma02} transit model around that star averaged over the transit duration (using limb darkening parameters for each target star from \citet{claret2012}), $\sigma$ being the published median three-hour combined differential photometric precision (CDPP) for that star, $N_{\rm pts}$ being the the total transit duration $T_{\rm dur}$/(3 hr), and $N_{\rm tr}$ being the orbital period of the simulated planet divided by the total amount of time that star was observed by \Kepler\ (number of quarters up until Q12, $\times$ 90 days, excepting Q1, which is 33 days).  We note that while this formulation ignores details of the noise properties on the exact timescale of transit and the exact observing window function, this prescription for calculating SNR is exactly the use case of the CDPP values as recommended by the \Kepler\ team \citep{christiansen2012}, especially since most of the targets in this survey are faint and thus white-noise dominated on the timescale of transits.  2.6\% of the planets simulated in this exercise---50,000 for each target star---have transiting geometries; this is the survey- and period-averaged transit probability discussed in \S\ref{sec:occurrence}.

\begin{figure}[t!]
   \centering
   \includegraphics[width=3.5in]{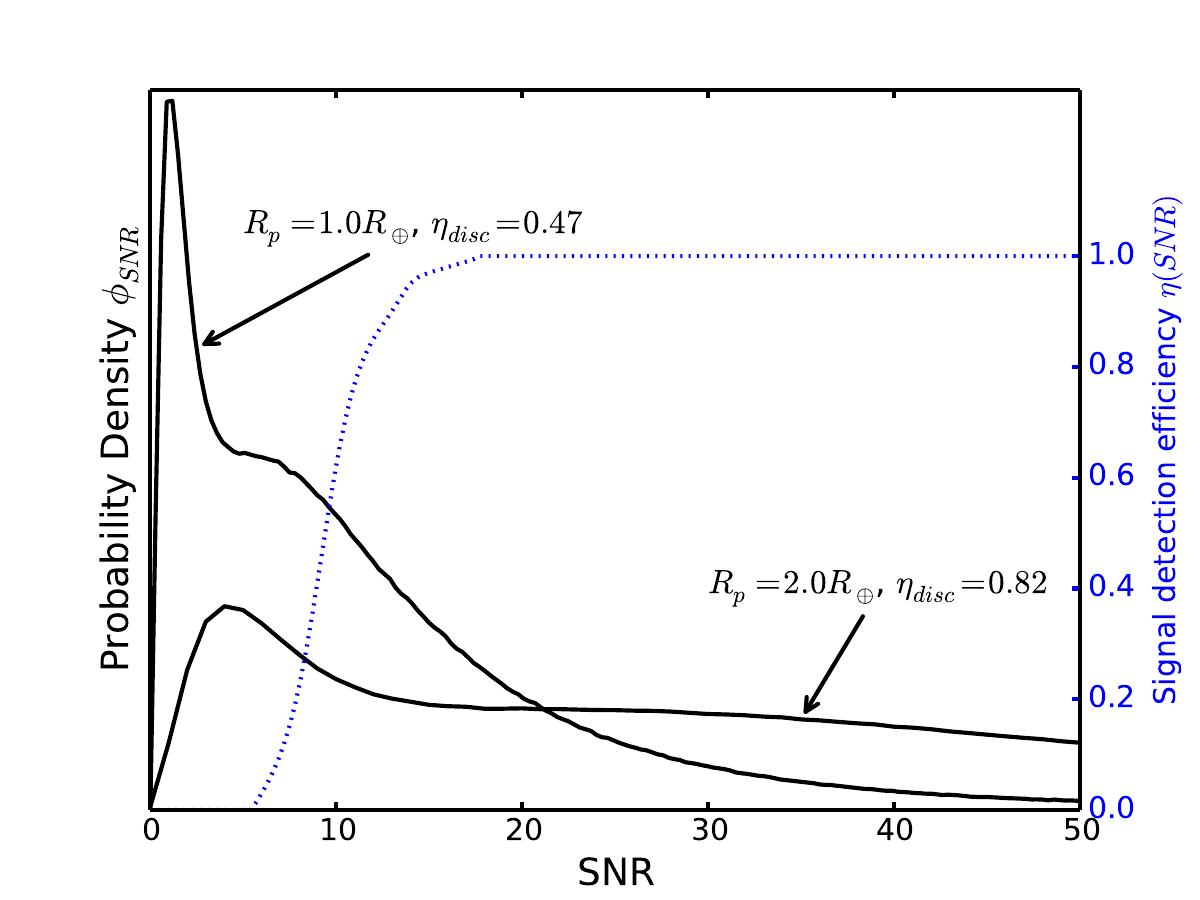} 
   \caption{Two examples of the SNR distributions resulting from simulating transits of a planet of a certain size around every \Kepler\ target star considered in this study, randomly assigning periods and impact parameters.  The properties of this distribution depend on the radii and photometric noise levels of the target stars, and the integral of the product of the pipeline detection efficiency function (blue dotted curve, from P.~Tennenbaum, priv.~comm.) with this distribution  gives the ``discovery fraction'' $\eta_{\rm disc}$ (Equation \ref{eq:etadisc}).  A 1 $R_\oplus$ planet is detectable in only about half of potential transiting configurations, whereas a 2 $R_\oplus$ planet is detectable in about 4/5 of potential configurations.}
   \label{fig:snrdists}
\end{figure}

\begin{figure}[t!]
   \centering
   \includegraphics[width=3.5in]{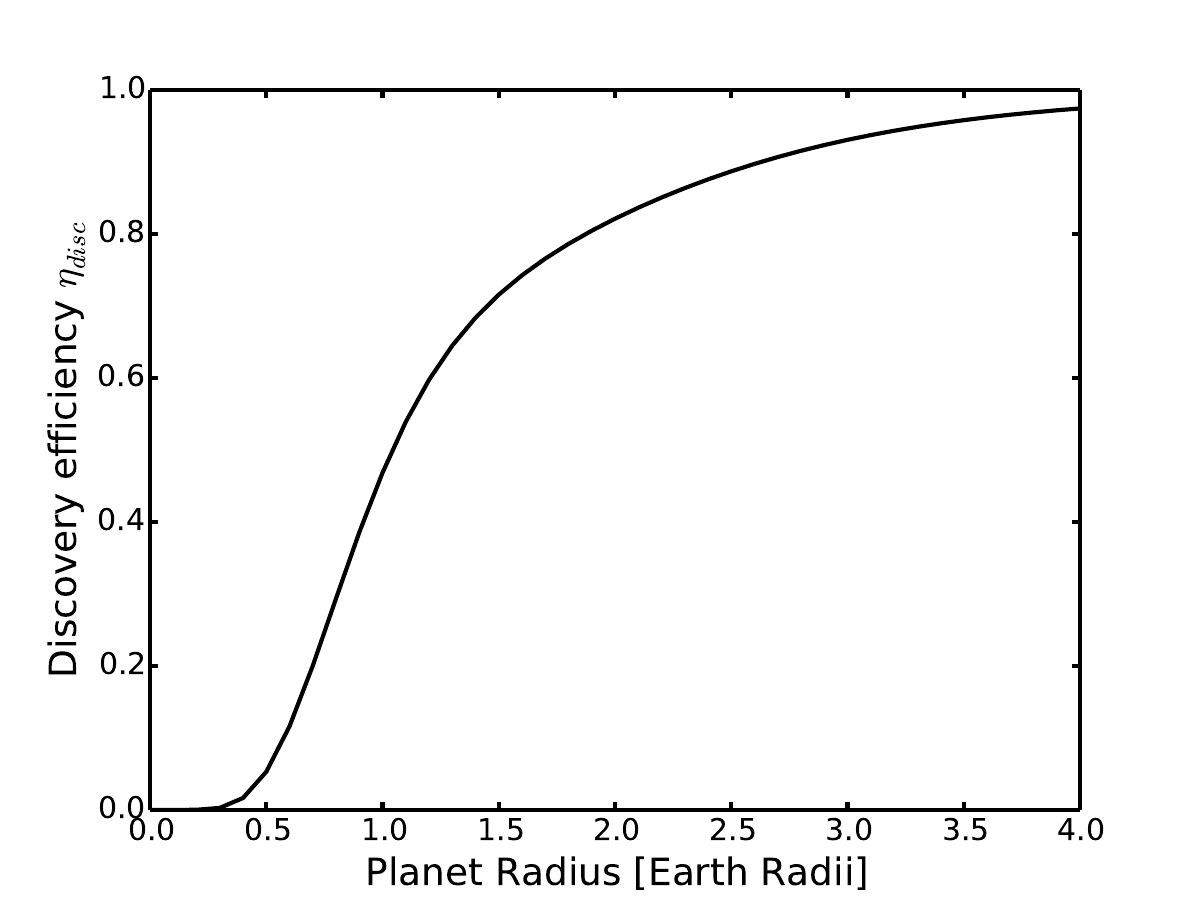} 
   \caption{The discovery efficiency of the \Kepler\ pipeline as a function of planet radius for the subset of \Kepler\ target stars considered in this work.  At each value of planet radius, $\eta_{\rm disc}$ is calculated according to Equation \ref{eq:etadisc}, using the SNR distributions as calculated in \S\ref{sec:snrdist} and illustrated in Figure \ref{fig:snrdists}.}
   \label{fig:etadisc}
\end{figure}

Rather than repeat this exact procedure for every planet, it is sufficient to generate the distribution once for a fiducial planet radius $r_0$ (e.g., 1 $R_\oplus$), and then create the distribution appropriate for any other planet with radius $r$ by multiplying all the SNRs in the fiducial distribution by $(r/r_0)^2$.  Figure \ref{fig:snrdists} illustrates these distributions for 1 $R_\oplus$ and 2 $R_\oplus$.  Additionally, $\eta_{\rm disc}$ as a function of planet radius $r$ may be determined by simply evaluating Equation \ref{eq:etadisc} along a grid of SNR distributions corresponding to an evenly spaced grid in planet radius.  Figure \ref{fig:etadisc} illustrates this result: discovery completeness rises from zero just below $r = 0.5$ $R_\oplus$ to nearly unity at $r = 4$ $\Rearth$.

\subsection{Planet Radii}
\label{sec:radii}
Before being able to apply the formalism discussed in \S\ref{sec:radfunction} to derive an estimate of the planet radius function, we must first understand the radii of the detected planets.  The radius of a transiting planet is extracted from its transit light curve, the depth of which reveals---approximately---the planet-star radius ratio $r_p/R_\star$.  However, there are subtle degeneracies between the radius ratio and other parameters of the fit (impact parameter in particular), especially when the duration of the transit begins to be comparable to the photometric integration time.   The \Kepler\ long-cadence integration time is 29.4 minutes, and since many of the stars in this particular study are relatively small, the durations of the transits are often only 2-3 times this.  Consequently, the shapes of the transits are artificially smoothed, appearing more V-shaped than they would be with shorter-cadence data, and leading to even greater degeneracy between the planet/star radius ratio and the transit impact parameter.

Thus, in order to understand each planet radius, the NExScI archive catalog values (derived from the \Kepler\ pipeline maximum-likelihood fits) are insufficient: a Markov Chain Monte Carlo (MCMC) approach is required.  Our procedure for extracting, detrending and fitting the transit signals will be described in detail in a forthcoming publication (Swift et al. in prep), but we briefly summarize the process here. Pre-search Data Conditioning Simple Aperture Photometry (PDCSAP) light curves are used from the 23rd data release for our analyses. Transit signals are selected from the light curves based on the catalog values for planet period and duration. Data within 1.5 times the duration of the transit center is preserved and the out of transit data is detrended using a linear model. We then use the \texttt{emcee} python module \citep{emcee} to sample the posterior probability distributions for the parameters in our transit model based on \citet{ma02} using a quadratic stellar limb darkening with coefficients taken from \citet{claret2012}.

To convert the posterior PDFs for $r_p/R_*$ of each transit signal to a posterior on $r_p$, we do a joint Monte Carlo sampling from the $r_p/R_\star$ distribution and a distribution for $R_\star$ according to the derived values from either spectroscopic studies (where available) or \citet{dressing2013}.

\subsection{False Positive Probabilities}
\label{sec:fpp}

A transiting planet candidate may be an astrophysical false positive rather than a  \textit{bona fide} planet, as has been discussed extensively in literature regarding the \Kepler\ survey.  In this work, we calculate the false positive probabilities (FPPs) of each of the Cool KOI candidates using the procedure described in \citet{morton2012}, supplemented by the dispositions on the NExScI archive (for dispositions of ``FALSE POSITIVE'' we assign FPP=1).  Calculating FPP for a transit candidate inevitably requires a prior assumption about planet occurrence; in \citet{morton2012} this is quantified by the ``specific planet occurrence rate'' $f_{p,i}$, which is an assumed occurrence rate of planets between 2/3 and 4/3 of the radius of planet $i$.  

As the goal of the current study is determining exactly the quantity that is used for the planet prior in the FPP calculations (planet occurrence as a function of radius), and since the radius function study itself depends on FPP, there is an opportunity for  iterative convergence.  In other words, we may start with initial (not necessarily carefully calculated) values for $f_{p,i}$, determine the radius distribution $\hat\phi_r^{\Pmax}$ (as described in the following subsection) using the FPP values implied by this assumption, then recalculate $f_{p,i}$ by integrating over the derived radius distribution.  If these ``post-calculated'' $f_{p,i}$ do not match the initial values used, then the FPPs can be recalculated using these new $f_{p,i}$ values, and the process repeated until the output $f_{p,i}$ match the input.  The final result of this process is that of the 130 KOIs in this sample, 11 have FPP $>$ 0.9 (and are thus essentially ignored in generating the radius distribution) and 95 have FPP $<$ 0.10.  FPPs for individual KOIs are listed in Table \ref{table:results}.  We note that these FPP calculations are based on the \Kepler\ pipeline-derived $r/R_\star$ values, so for KOIs in this sample whose MCMC-derived radii are more than 30\% discrepant from the \Kepler\ pipeline-derived values, we manually set a lower limit to the FPP of 0.1; these KOIs (15) are indicated in Table \ref{table:results} as well.

\subsection{Radius Distribution}
\label{sec:rdist}

\begin{figure*}[htbp]
   \centering
   \includegraphics[width=\textwidth]{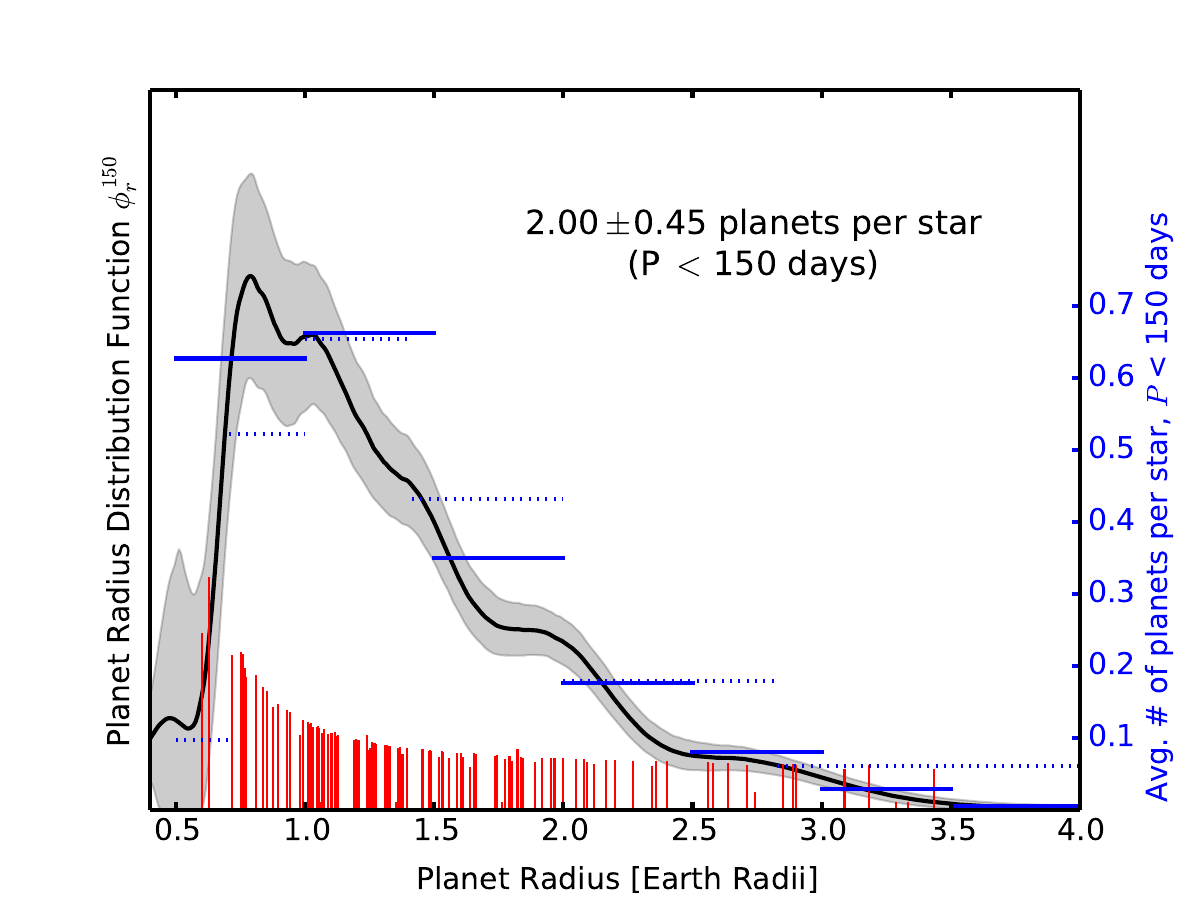} 
   \caption{The empirical radius distribution of planets orbiting M dwarfs with periods $<$\Pmax\ days (black continuous curve), estimated with a weighted kernel density estimator (wKDE; see \S\ref{sec:rdist}), with the bootstrap resampling-derived 1$\sigma$ uncertainty swath shaded grey---analogous to a running poisson error bar.  The detection efficiency as a function of signal-to-noise ratio has been quantified according to the estimate from the \Kepler\ detection pipeline shown in Figure \ref{fig:snrdists}.  The blue horizontal lines represent the standard ``occurrence rate per bin'' calculations for this sample, derived by integrating the density estimator over each bin.  The solid blue lines are linearly-spaced bins; the dotted blue lines are the logarithmically spaced bins used by \citet{dressing2013} and \citet{petigura2013b}.  The vertical red lines represent the radii of individual planets in the sample, with their heights being proportional to the weight factors $w_i$.   There is an average of $\npps \pm \enpps$  planets per cool star in orbits $<$\Pmax\ days over this radius range, and there is an average of greater than 1 planet per cool star in this period range for radii between 0.5 and 1.5 $\Rearth$.}
   \label{fig:rdist}
\end{figure*}

With the $r_p$ posterior distribution $p_{r,i}(r)$ for each KOI in hand, we then construct the wKDE to estimate the planet radius function.  The discovery efficiency $\eta_{{\rm disc},i}$ for each planet is given by integrating the curve in Figure \ref{fig:etadisc} over $p_{r,i}(r)$, and multiplied with the average transit probability of 0.026 gives the total efficiency factor $\eta_i$.  We then calculate FPP values using the iterative method described in \S\ref{sec:fpp}, which combined with $\eta_i$ gives $w_i$, according to Equation \ref{eq:weightsimple}, and subsequently the radius function $\hat\phi_r^{\Pmax}$.  We choose the  smoothing kernel for each planet to be $p_{r,i}(r)$, without any additional smoothing, and de-bias the estimator and calculate its variance according to the procedure described in \S\ref{sec:biasvariance}.  Figure \ref{fig:rdist} illustrates the result, with the 1$\sigma$ uncertainty region as the grey shaded region.  Table \ref{table:results} lists the planet radii with uncertainties, discovery efficiencies, and FPPs used to generate this radius distribution.

\begin{deluxetable}{ccccccc}
	\tabletypesize{\footnotesize}
	\tablecolumns{7}
	\tablecaption{Data Used in Radius wKDE}
	\tablehead{ \colhead{KOI} &\colhead{$R_{p,\oplus}$} & \colhead{$+1\sigma$} &  \colhead{$-1\sigma$} &  \colhead{$\eta_{\rm disc}$} & \colhead{FPP} & \colhead{$w_i$}}
	\startdata
2842.03\tablenotemark{a} & 0.60 & 0.24 & 0.18 & 0.21 & 0.1\tablenotemark{c} & 162.6\\
961.03\tablenotemark{a} & 0.63 & 0.16 & 0.15 & 0.16 & 0.1\tablenotemark{c} & 214.7\\
4252.01\tablenotemark{a} & 0.72 & 0.15 & 0.09 & 0.26 & 0.018 & 143.0\\
2662.01\tablenotemark{a} & 0.75 & 0.10 & 0.07 & 0.26 & 0.0041 & 145.1\\
2842.01\tablenotemark{a} & 0.76 & 0.22 & 0.18 & 0.29 & 0.1\tablenotemark{c} & 121.0\\
961.01\tablenotemark{a} & 0.76 & 0.18 & 0.18 & 0.26 & 0.0091 & 143.9\\
1843.02\tablenotemark{a} & 0.76 & 0.10 & 0.07 & 0.28 & 0.01 & 138.3\\
2542.01\tablenotemark{a} & 0.77 & 0.13 & 0.12 & 0.28 & 0.069 & 130.1\\
255.01\tablenotemark{a} & 0.77 & 0.19 & 0.12 & 0.33 & 0.1\tablenotemark{c} & 105.3\\
4875.01\tablenotemark{a} & 0.77 & 0.17 & 0.13 & 0.31 & 0.021 & 122.1\\
: & : & : & : & : & : & : \\
251.01\tablenotemark{a} & 2.89 & 0.21 & 0.22 & 0.92 & 0.0017 & 41.7\\
250.02\tablenotemark{a} & 2.90 & 0.29 & 0.34 & 0.92 & 0.02 & 41.0\\
2793.01\tablenotemark{b} & 3.09 & 0.43 & 0.38 & 0.93 & 0.1\tablenotemark{c} & 37.1\\
250.01\tablenotemark{a} & 3.18 & 0.30 & 0.30 & 0.94 & 0.011 & 40.4\\
1006.01\tablenotemark{b} & 3.29 & 0.50 & 0.39 & 0.95 & 1\tablenotemark{c} & 0.0\\
886.01\tablenotemark{a} & 3.33 & 11.09 & 1.95 & 0.76 & 1 & 0.0\\
1879.01\tablenotemark{a} & 3.43 & 0.58 & 0.53 & 0.95 & 0.096 & 36.6\\
531.01\tablenotemark{a} & 4.14 & 0.47 & 0.37 & 0.98 & 0.99 & 0.3\\
1681.01\tablenotemark{a} & 4.69 & 19.40 & 3.19 & 0.68 & 0.99 & 0.3\\
2992.01\tablenotemark{b} & 5.18 & 31.57 & 3.52 & 0.59 & 0.88 & 7.7
	\enddata

	\tablenotetext{a}{Planet radius based on spectroscopic stellar parameters from the analysis of \citet{muirhead2012b} or Muirhead et al. (in prep).}
	\tablenotetext{b}{Spectroscopic stellar characterization not available, so planet radius based on stellar parameters from \citet{dressing2013}}
	\tablenotetext{c}{FPPs for these KOIs are not calculated. For those that are dispositioned ``FALSE POSITIVE'' on the NExScI archive, a value of 1 is assigned; for those for which the \Kepler\ pipeline-derived planet radii are more than 30\% discrepant from the $R_p$ listed here, a FPP of 0.10 is assigned.}
	\label{table:results}
\end{deluxetable}

\section{Results}
\label{sec:results}

The overall normalization of the estimated radius function shown in Figure \ref{fig:rdist} indicates that there are $\npps \pm \enpps$ planets per cool star with periods $<$150d.  The most notable feature of this distribution is that it rises more or less smoothly with decreasing radius down to below 1 $\Rearth$.  The estimator then appears to decrease again for radii smaller than $\sim$0.8 $\Rearth$. While this turnover may indeed be real, we show in \S\ref{sec:validation} that such a feature may be present even if the true underlying distribution continues to rise.

If the turnover is indeed a robust feature of the distribution, its explanation might be similar to our current understanding of the origins of the inner Solar System \citep{goldreich2004,chambers2001}: a large number of isolation-mass protoplanets form quickly, and once the gas and planetesimal disk dissipates, a period of dynamical instability follows, at the end of which typically only a few larger planets remain, the rest having been either destroyed (or merged) via collisions or swallowed by the host star.  It is certainly plausible that $\sim$1 $\Rearth$ planets might generically be the most likely outcome of this process, as this is precisely what has happened with the inner Solar System, with an outcome of two planets about the size of Earth. 

This distribution also indicates that planets larger than $\sim$3 $\Rearth$ are very rare around cool stars, consistent with the findings of RV surveys \citep{endl2003,johnson2010a,johnson2010c,bonfils2013}.   There has been one hot Jupiter identified around a star in this sample \citep[KOI-254b/Kepler-45b][]{johnson2012} and another recent discovery of note \citep{triaud2013}, but such planets are clearly exceptional---the vast majority of close-in planets around cool stars are smaller than $\sim$3 $\Rearth$.  Even Gliese 1214b \citep{charbonneau2009}, by far the best-studied planet around an M dwarf to date, appears to be an exception to the typical system, as its radius of 2.7 $\Rearth$ lies far down the tail of this distribution.  In fact, there are $\sim$20$\times$ more planets smaller than Gl 1214b than there are larger than Gl 1214b---this bodes very well for the future of ground-based surveys, both transit and RV, as they become more sensitive to smaller planets.

We also note that Figure \ref{fig:rdist} illustrates how the qualitative interpretation of the radius function changes significantly moving from histogram presentation to the non-parametric density estimator.  Overlaid on the the $\phi_r^{150}$ function plotted in Figure \ref{fig:rdist} are histograms of planet occurrence rate in two different sets of bins---the solid blue lines being linearly spaced bins and the dotted lines being logarithmically spaced bins, the binning choice for most \Kepler\ occurrence analyses to date \cite{howard2012,dong2012,dressing2013,fressin2013,petigura2013, petigura2013b}.  

The qualitative understanding of planet occurrence patterns communicated by these two different binning schemes is quite different, and both obscure the true underlying detail of the distribution.  A quick glance at the occurrence in logarithmic bins gives an impression that the peak of the planet radius distribution is between 1 and 1.5 $\Rearth$, perhaps around 1.25 $\Rearth$, while the true peak is really below 1 $\Rearth$.  The linearly binned histogram tells a different story: according to this binning, the impression is that planet occurrence rises with decreasing radius until 1.5 $\Rearth$ and then is constant from there down to 0.5 $\Rearth$.  This characterization is also quite inaccurate, as the planet radius distribution function at 0.8 $\Rearth$ is actually nearly twice that at 1.5 $\Rearth$.  Understanding the shape of the planet radius distribution in detail inaccessible to histogram binning is especially important since it is known that the physical properties of planets likely change dramatically between $\sim$1 and $\sim$2 $\Rearth$ \citep{weiss2014,marcy2014,lopez2013}.

\section{Discussion}
\label{sec:discussion}

In this section, we first discuss the assumption we have made that the period distribution of planets is independent of radius.  We then explore differences between the methods presented in this paper and those more commonly used in the literature to derive the planet radius distribution.  Finally, we validate that the approach used in this work more accurately recovers the true radius function than does the more commonly used approach.

\subsection{Period Distribution Assumption}
\label{sec:perdistassumption}

Figure \ref{fig:pincomplete} illustrates very clearly why the detected population of small planets in the Cool KOI sample is indeed very likely incomplete, showing that where the detected period distribution of the smallest of the Cool KOIs drops off is right around the periods where the known short-period small KOIs would have become undetectable.  This is the motivation behind the period redistribution procedure we use to calculate $\eta_{{\rm disc},i}$ in \S\ref{sec:calculation}---correcting for the undetectable longer-period small planets.  However, the nature of this correction as applied in this work---using the implied all-planet period distribution for each planet---merits some discussion.

\begin{figure}[t!]
   \centering
   \includegraphics[width=3.5in]{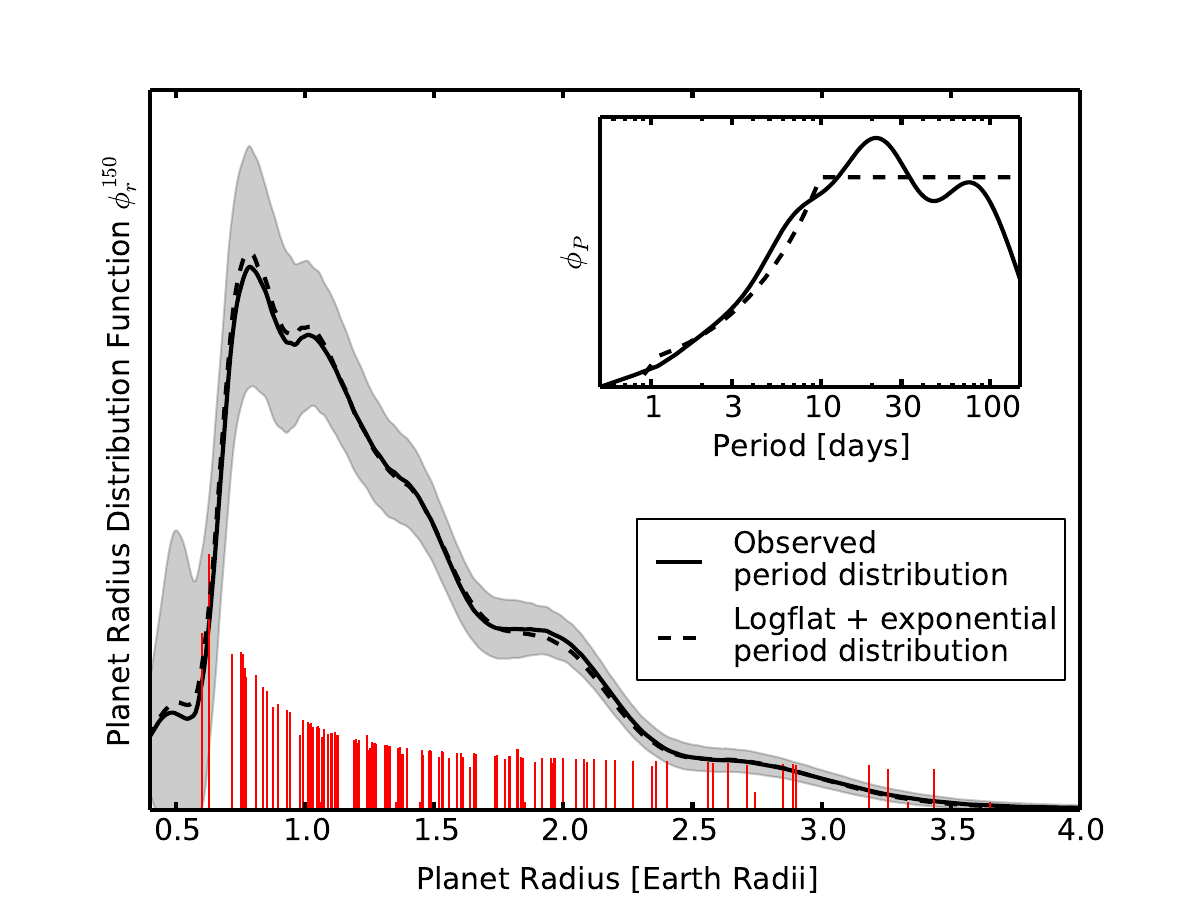} 
   \caption{Comparing the radius function derived in this work to one derived assuming the period distribution of all planets is as the dashed line in the inset figure, rather than the sold line.  The difference in the resulting distribution is negligible; within the uncertainties intrinsic to the de-biasing procedure.}
   \label{fig:rdistcomparepdist}
\end{figure}

There are certainly both physical reasons and observational suggestions to believe that the planet period distribution is \emph{not} completely independent of radius.  In particular, \citet{howard2012} finds (shown in their Figure 6) that the fraction of short-period planets that are large (4-8 $\Rearth$) is smaller than the fraction of longer-period planets that are large; in other words, the period distribution of larger planets decreases (heading towards shorter periods) sooner than does the distribution of smaller planets (2-4 $\Rearth$).  \citet{dong2012} present a similar finding.  While there is not yet compelling evidence that this same effect has been detected for planets smaller than 2 $\Rearth$, simple physical considerations such as increasing stellar insolation \citep{weiss2013} might reasonably contribute to a dearth of larger planets on short-period orbits.   However, as there is no corresponding clear physical explanation for the absence of smaller planets in longer orbits, it is reasonable to assume that they do in fact exist, and that their period distribution might resemble the period distribution of the larger planets that are detected in such orbits.

In order to explore this in detail, we repeat the analysis assuming a modified period distribution, shown in the inset of Figure \ref{fig:rdistcomparepdist}.  Rather than using a wKDE built from the observed periods of the detected planets, we use a log-flat distribution for periods greater than 10 days, with an exponential cutoff short of 10 days.  Qualitatively, this seems to approximately match the observed distribution, except that instead of tailing off towards longer periods it remains constant.  The radius function that results from assuming this period distribution is plotted as the dashed line in the main panel of Figure \ref{fig:rdistcomparepdist}, and differs only very slightly from the original analysis, within the noise inherent in the de-biasing procedure.  The reason for this negligible difference is that the bulk of the correction for small planets happens when it is assumed that the planets observed with $\lesssim$1d periods can also exist at periods of $\sim$10's of days---in other words, a 0.7 $\Rearth$ planet is just as undetectable at a period of 40 days as at 100 days.  We thus conclude that our results are not very sensitive to the details of the longer-period radius distribution, but rather depend only on the assumption that small planets do indeed exist at periods beyond which they are detectable in a manner roughly similar to that of larger planets.

\subsection{Comparison with other methods}
\label{sec:comparison}

The method we have used to calculate the period-marginalized radius function contains several significant differences from previous planet occurrence rate studies.  In particular, we use a survey- and period-averaged approach to calculating both the discovery efficiency of a planet and the transit probability correction factor, as opposed to the more traditionally used method of correcting each planet individually---that is, simulating the planet around other stars at only its observed period, and using each planet's individual transit probability.  We summarize this difference as calculating the radius function while ``pre-marginalizing'' over period.

\begin{figure}[t!]
   \centering
   \includegraphics[width=3.5in]{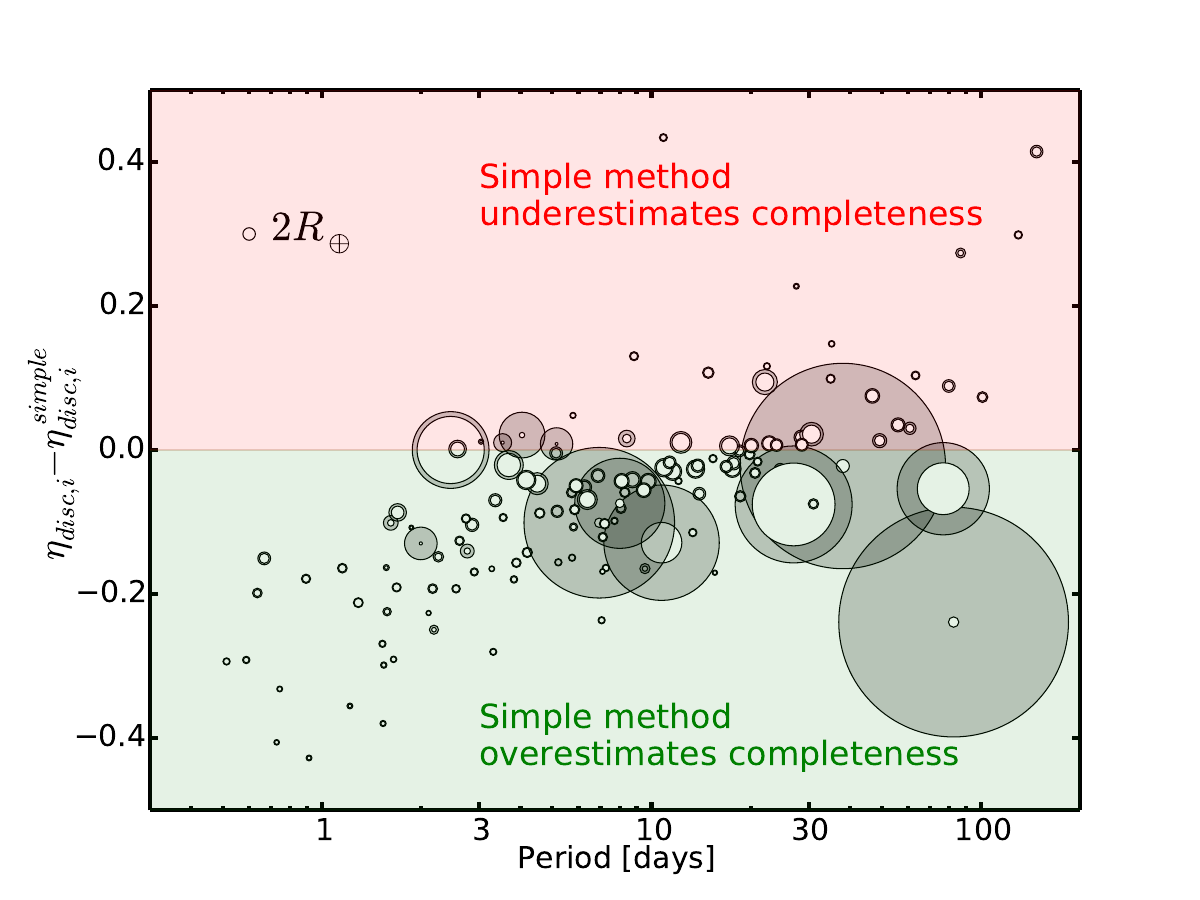} 
   \caption{The difference between the discovery fraction $\eta_{\rm disc}$, calculated by simulating ``alternative scenario'' planets around other stars with a distribution of periods, and $\eta_{\rm disc}^{\rm simple}$, calculated by simulating planets only at the discovered periods.  The sizes of the points are proportional to planet radius, with the annuli representing the $\pm$1$\sigma$ uncertainties from the MCMC posteriors.  The discovery fractions of smaller planets, which are found primarily at shorter periods, are typically underestimated by the simple method.}
   \label{fig:etacompare}
\end{figure}

Figure \ref{fig:etacompare} illustrates the difference in discovery efficiency calculated using the method used in this work ($\eta_{{\rm disc},i}$ as described in \S\ref{sec:calculation}) and that which would be calculated by only simulating planets around other stars at the single detected period: $\eta_{{\rm disc},i}^{\rm simple}$.  In this figure, the size of the points is proportional to planet radius, with the annuli representing the $\pm 1 \sigma$ uncertainties from the radius posteriors.  Predictably, planets discovered at short periods---predominantly small---get smaller $\eta_{{\rm disc},i}$ (larger correction factor) than the simple ``post-marginalized'' calculation; this is because the majority of the period-redistributed simulated planets will be at longer periods than the original, and thus have lower SNR.  Planets discovered at longer periods tend to have the opposite effect.

We also note that to get $\eta_{{\rm disc},i}$ in our calculations, we integrate the posterior distribution of planet radius over the efficiency curve in Figure \ref{fig:etadisc}; for $\eta_{{\rm disc},i}^{\rm simple}$, we simulate the alternative scenarios as having the planet radius fixed to be the median of its posterior---this explains the scatter of this difference around a monotonic relationship to period.  Additionally, the efficiency curve we use is not just a simple SNR threshold, as has been used in some previous studies.

In addition, in order to correct for the planets missed by \Kepler\ due to random orbital orientations, we do not correct each planet individually for its observed transit probability; rather we use the same single survey- and period-averaged transit probability for every planet.  The justification for this is that only way to properly determine $\eta_i$ is through the full Monte Carlo procedure discussed in \S\ref{sec:calculation}, where planets are simulated isotropically, the SNR distribution of the ones with transiting orientations are used to calculate $\eta_{{\rm disc},i}$, and the fraction  \textit{of the whole simulation} that transits is used as $\eta_{{\rm tr},i}$.  Alternatively, one could imagine ignoring the factorization entirely and just assigning SNR = 0 (with detectability = 0) to the non-transiting planets in order to directly calculated $\eta_i$, and the result would be the same (though less intuitive).  

\begin{figure}[t!]
   \centering
   \includegraphics[width=3.5in]{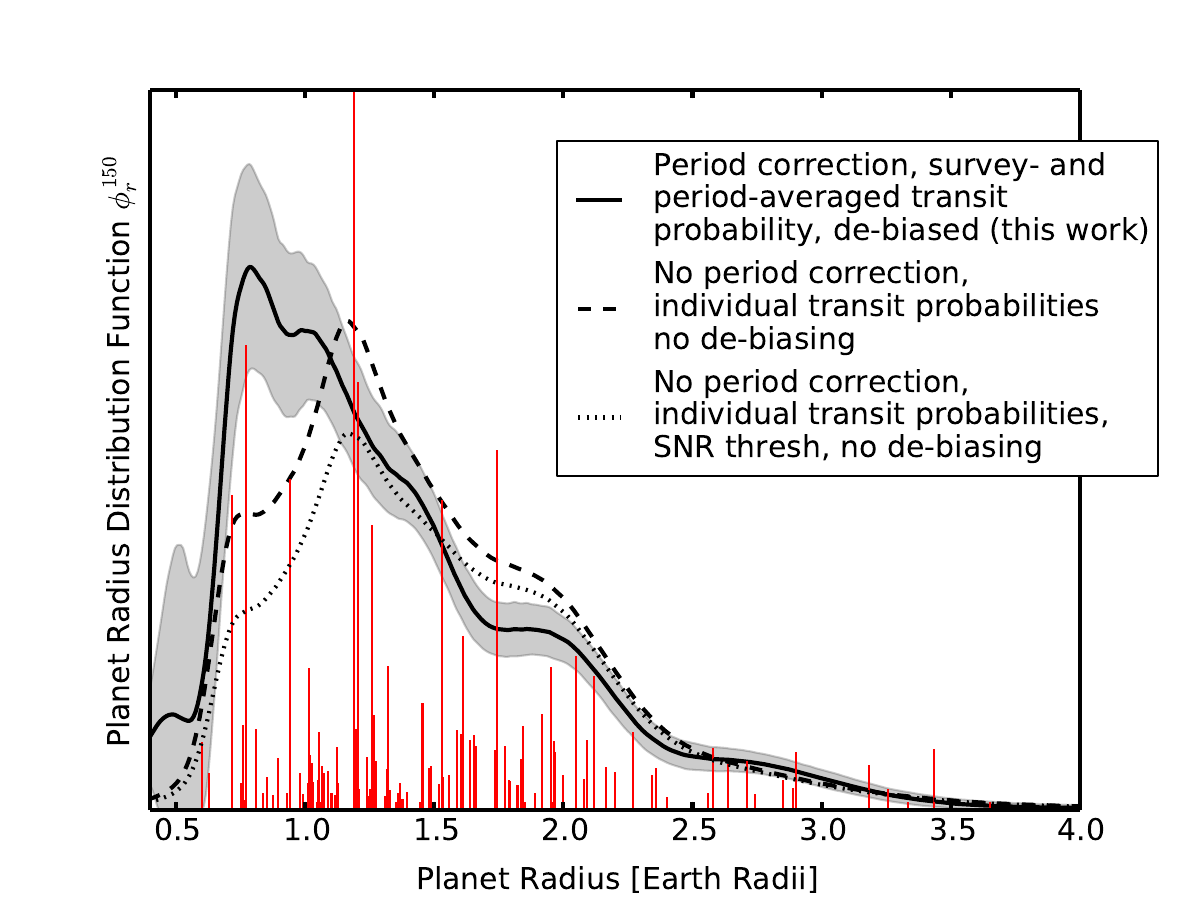} 
   \caption{Comparing the radius function calculated in this paper (solid line with grey uncertainty region) with the radius distributions that would be derived by calculating completeness corrections according to the methods typically used in the literature.  Both the dashed and dotted lines correct for transit probability on an individual basis---i.e., each weight factor $w_i$ is proportional to the inverse of the geometric transit probability of planet $i$, rather than using the survey- and period-averaged transit probability we advocate.  The dashed line uses the same discovery efficiency as a function of SNR as we do in this work (Figure \ref{fig:etadisc}), whereas the dotted line uses a sharp threshold cutoff at SNR = 7.1.  The heights of the red lines are proportional to the weights $w_i$ under the dashed-line calculation.  Neither the dashed nor dotted lines implement the de-biasing procedure, which we introduce in this work.  We see that using individual transit probabilities moves the peak of the radius function to about 0.25 $\Rearth$ larger than it would otherwise be; in addition, using a SNR threshold significantly decreases the assumed occurrence rate of smaller planets.}
   \label{fig:rdistcompare}
\end{figure}

Figure \ref{fig:rdistcompare} illustrates how the results of this work would differ if we used the same data, but did our calculations according to the more widely used methods of completeness correction.  The dotted line illustrates a calculation in the style of \citet{dressing2013} or \citet{howard2012}.  Each planet is corrected individually for its own transit probability rather than using a global average, and the detection efficiency curve is taken to be a step function at SNR = 7.1.  We see that the occurrence rate of small planets is significantly underestimated.  The dashed line is an improved version of this calculation, using the detection efficiency as a function of SNR from Figure \ref{fig:etadisc}, but still correcting for transit geometry using only individual transit probabilities and without period redistribution in the $\eta_{\rm disc}$ calculation (this is analogous to the method used by \citet{petigura2013b}).  We do not de-bias either the dotted or dashed distributions.  The overall normalization is lower than our calculations (solid line) and the qualitative shape of the radius distribution has shifted, showing a peak around 1.2 $\Rearth$ and a clear decline below, rather than the continued rise to below 1 $\Rearth$ that we find.  

This qualitative reason for this discrepancy is quite simple:  the dashed and dotted lines in Figure \ref{fig:rdistcompare} are plotted down to below the planet radius where the survey is able to reliably detect planets on 150-day orbits (in this case, about 1 $\Rearth$).  Whereas the method used in this paper ``pre-marginalizes'' over period, enabling extrapolation of the smaller-radius population to longer periods, the ``simple'' individual method does not do this.  The simple method implies ``post-marginalization,'' which can only be valid if the survey is actually sensitive enough to planets in the longest-period and smallest-radius corner of the period-radius space under consideration.  So the simple method is not wrong if properly applied, but it cannot be used to estimate the radius function down to as low a radius as the method advocated in this paper.  Additionally, it is easy to see from Figure \ref{fig:rdistcompare} how the shape of the simple-method estimator is much more sensitive to random fluctuations in the data (due to the large correction factors for long-period planets) than the pre-marginalized estimator; this can also cause spurious peaks in the distribution that are strongly dependent on the particular realization of the data that we happen to observe.

\subsection{Validation of methods, and the limitations of radius function estimation}
\label{sec:validation}

\begin{figure}[t!]
   \centering
   \includegraphics[width=3.5in]{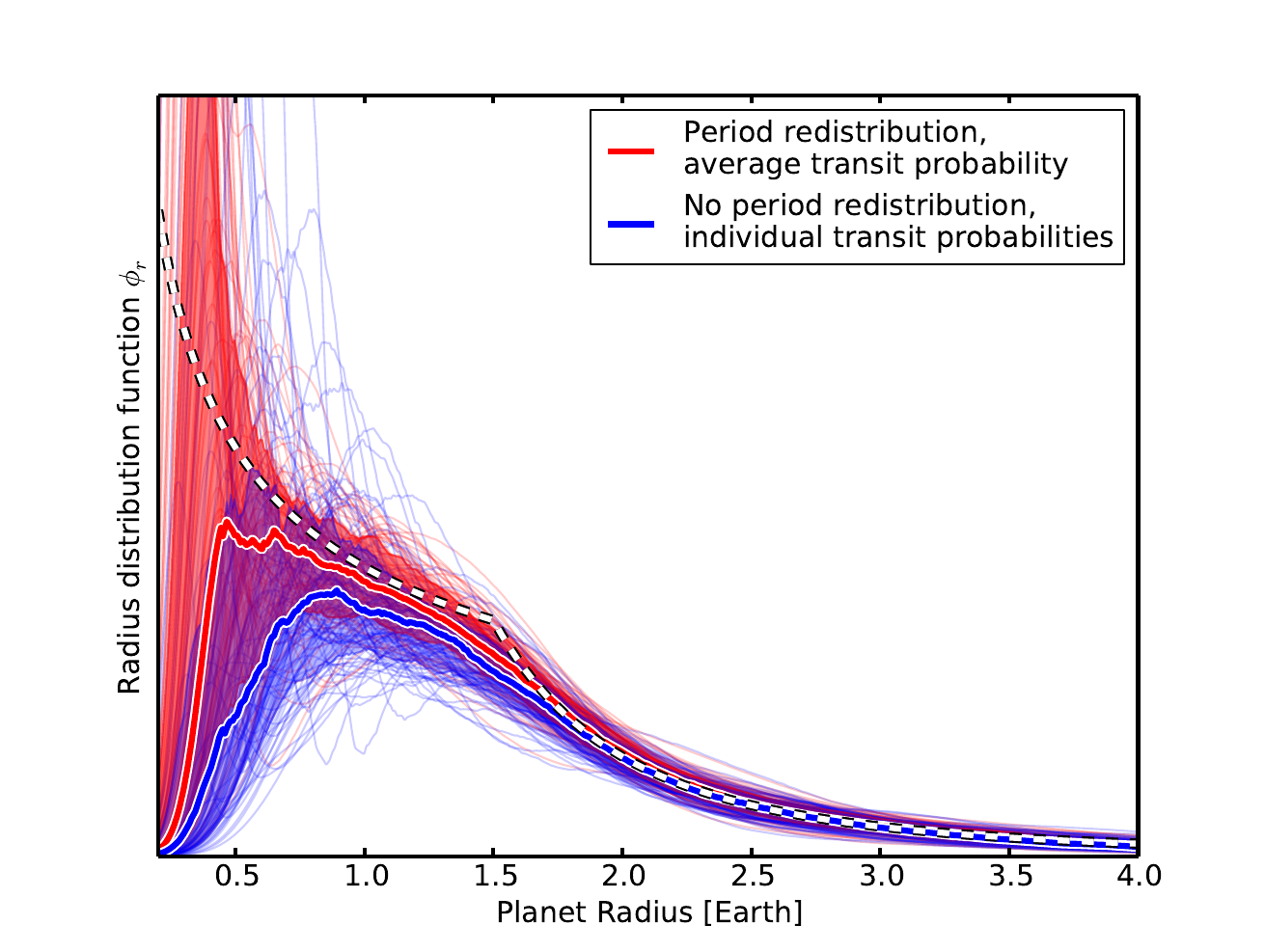} 
   \caption{A validation test of the wKDE method presented in this work, and comparison of two different completeness-correction recipes.  We generate 100 mock transit survey datasets, using the same target stars and signal detection criteria as in this study, with planet radii drawn from the distribution illustrated by the dashed white line.  Each thin line is a de-biased wKDE radius function derived according to the methodology presented in this paper.  Two different methods are used to calculate the weights for the wKDEs---red lines use the period-averaging strategy we advocate in this paper; blue lines use the more standard procedure of using individually calculated transit probabilities and calculating the discovery efficiency by simulating planets around other stars by fixing the observed period.  Heavy solid lines are the median of the ensemble of distributions for each color, and the colored swaths indicate the $\pm$1$\sigma$ percentile range of the thin lines.  Both methods recreate the distribution nearly exactly for $r_p > 1.5$ $\Rearth$; for $r_p < 1$ $\Rearth$, the period-averaging method performs significantly better.  Notably, the blue method typically predicts a turnover of the radius distribution around 1 $\Rearth$, even though the true distribution continues to rise.  Conversely, the typical red reconstruction does indeed capture a continued rise down to 0.5 $\Rearth$, where the detection efficiency is only 5\%.  However, the variance of this estimator is quite large at these small radii, and many of the data realizations in both methods result in a turnover.  Therefore, this experiment demonstrates that detection of a turnover in the reconstructed radius distribution does not rule out a continued rise in the true underlying distribution.}
   \label{fig:methodcompare}
\end{figure}

While we have demonstrated and discussed differences between the planet radius function derived using two different techniques of calculating individual weights, the question still remains whether or how well either can actually accurately reconstruct the true radius distribution.  

To explore this, we run a Monte Carlo experiment where we first generate 100 different transit survey data sets with planet frequency and radii drawn from a known distribution---a broken power law normalized to 3 planets/star on the interval [0.3,4.0] $\Rearth$, proportional to $r_p^{0.5}$ for  $r_p < 1.5$ $\Rearth$ and to $r_p^{-3}$ for $r_p > 1.5$ $\Rearth$ (the same distribution used in the de-biasing demonstration in Figure \ref{fig:debiasing}).  The periods of these simulated planets are assigned according to the log-flat/exponential distribution illustrated in Figure \ref{fig:rdistcomparepdist}.  For each of these mock data sets, we derive and de-bias the wKDE estimator $\hat\phi_r^{150}$ two different ways: first calculating weights $w_i$ according to the procedure described in \S\ref{sec:calculation}, and second calculating weights according to the ``simple'' prescription, without period redistribution and using individual transit probabilities.  

Figure \ref{fig:methodcompare} illustrates the results of this experiment.   Each thin red line is a de-biased wKDE derived using the methods presented in this paper to calculate detection efficiencies; each thin blue line is a de-biased wKDE using the individual period/individual transit probability (simple) method.  The thicker lines and colored bands indicate the respective median and $\pm 1$$\sigma$ ranges from this ensemble. The white dashed line is the true underlying distribution from which the planet radii are drawn.  

There are several important points this experiment demonstrates.  First, both methods correctly recover the underlying distribution nearly perfectly for radii larger than 1.5 $\Rearth$, validating the accuracy of the technique.  Below 1.5 $\Rearth$, the two methods begin to diverge, with ours sticking closely to the true distribution until $r_p \lesssim 0.5$ $\Rearth$, but the simple method beginning to significantly underestimate planet occurrence by $r_p \lesssim $1 $\Rearth$.  Indeed, the typical result of the simple method is an estimate of a turnover in the radius function at $\sim$1 $\Rearth$---a strong qualitative discrepancy from the true distribution.   

We also note that at the very low end of the distribution ($r_p \lesssim 0.5$ $\Rearth$), neither method correctly reconstructs the continued rise, though the period-redistribution method does still keep the true distribution within the 1$\sigma$ percentile  range while the simple method does not.  This may be understood by realizing that if a very small planet does happen to be detected (which will only occur in rare cases since the detection efficiency is only $\sim$5\% at $r_p=0.5$ $\Rearth$), it will necessarily be at a very short orbital period, so the weight factor it receives without period correction will be strongly underestimated.   This confirms our qualitative understanding of the difference between the two methods---the post-marginalized simple strategy can only be valid down to a radius where the survey is decently complete at the maximum allowed period (this is apparently somewhere between 1 and 1.5 $\Rearth$), whereas the pre-marginalized method enables confident extrapolation to significantly below this point.

Perhaps most dramatically, Figure \ref{fig:methodcompare} illustrates that even if the true distribution continues to rise down to arbitrarily small radii, the estimator for its shape will typically turn over significantly above the actual detection limit of the survey, and begin to flatten at even larger radius.  In other words, our derived radius distribution from actual \Kepler\ data that we show in Figure \ref{fig:rdist} could very plausibly reflect a true underlying distribution that keeps rising continuously down to below 0.5 $\Rearth$---\textit{despite the fact that a log-binned histogram of the estimated distribution looks like it turns over below 1.5 $\Rearth$.}  

Finally, we note that the analysis of \citet{petigura2013b} has found, in apparent contradiction with our results, that planets in the $2.0-2.8$ $\Rearth$ radius bin are more common than planets in the $1.4-2.0$ $\Rearth$ bin, which are in turn more common than in the $1.0-1.4$ $\Rearth$ bin.  There are several possible explanations for this discrepancy.  First, the bins in question are logarithmically spaced, which immediately exaggerates the presence of any turnover, though this alone is not sufficient to explain the difference in results.  Second, while the bins used by \citet{petigura2013b} are chosen to all have significant numbers of detected planets, the completeness does vary from $\sim$70\% to $\sim$10\% within the smallest, longest-period bin used for the radius function reconstruction, potentially leaving room for some of the effects that distinguish the blue curve from the red in Figure \ref{fig:methodcompare}.  Another contributing explanation may also be that the only planets counted in \citet{petigura2013b} were the most-detectable planets in each system, which will lead to preferential underestimation of occurrence in the smallest radius bins.  Or most simply, this difference may just be a function of the stellar target sample considered---\citet{petigura2013b} used only Solar-like (G/K) stars in their target sample, whereas we use only stars with $T_{\rm eff} < 4000$ K.  If this is the reason, further study of \Kepler\ results should distinguish a radius function that changes shape with increasing host-star temperature.

\section{Conclusions}
\label{sec:conclusion}
 
We introduce and validate a simple non-parametric method of analyzing the empirical shape of the period-marginalized planet radius distribution from a transit survey---the weighted kernel density estimator, or wKDE.  This estimator is similar to a standard kernel density estimator, except that its overall normalization is constructed to be equal to the total number of planets per star, and that each data point is weighted according to its inverse detection efficiency.   While the naive construction of this estimator is most generally biased, we present a bootstrap-based method for de-biasing it, adapted from the ``smoothed bootstrap'' presented in \citet{narsky2013}.  We also show that the detection efficiency is best computed with a ``pre-marginalization'' procedure: re-distributing planets at all possible periods when calculating how many target stars around which a particular planet could have been observed, and using a survey- and period-averaged transit probability.  Additionally, we demonstrate that it is important to use a realistic detection efficiency as a rising function of signal-to-noise ratio, rather than a strict cutoff, to correctly calculate the occurrence rates of smaller planets.  We also emphasize that presentation of the planet radius function in histograms---especially with logarithmically spaced bins---runs a risk of qualitatively misrepresenting the true shape of the distribution.

Applying this analysis to the 130 planet candidates in the Q1-Q12 KOI catalog with periods less than \Pmax\ days discovered around the cool ($T_{\rm eff} < 4000$ K) \Kepler\ targets photometrically characterized by \citet{dressing2013}, incorporating their individually calculated false positive probabilities following \citet{morton2012} as well as new MCMC fits for planet radii (Swift et al., in prep), we find that the planet distribution continues to rise continuously down to at least $\sim$1 $\Rearth$ and possibly below.  We detect a possible turnover in the radius distribution below $\sim$0.8 $\Rearth$ but we demonstrate that it is very plausible that this may reflect an underlying distribution that continues to rise.  

This is the first radius-function reconstruction study that is sensitive to this range of planetary radius, and it appears to contradict the results of \citet{petigura2013b}, who use a ``post-marginalization'' completeness correction method and see a peak in the radius function around $\sim$2 $\Rearth$.  This contradiction may be due to either a confluence of small factors that could add up to a qualitative miscalculation in the \citet{petigura2013b} study, or it may simply be due to the different host star populations considered (G/K in that study, as opposed to M stars in this).  This certainly warrants further investigation, in particular investigating how the radius function may change with stellar type.  New detections by future transit missions such as K2, TESS, and PLATO will also provide additional insight into these questions.  

If the flattening/turnover of the radius function just below 1 $\Rearth$ is indeed a true feature of the distribution, it invites theoretical exploration, as it would suggest that planets around the size of Earth are the most common to survive the process of system formation and evolution around cool stars.  This outcome is certainly plausible, given the observed architecture of our Solar System.   And as a final note, the occurrence pattern of planets around cool stars indicates that there are indeed many planets just beyond the detection threshold of ground-based surveys, as planets larger than Gl 1214b (2.7 $\Rearth$) are $\sim$20$\times$ rarer than planets with $R_p < 2.7 \Rearth$.

Looking at the occurrence rates of the smallest planets in particular, we may compare with previous studies to estimate the degree to which the estimates of ``habitable-zone'' planets might change with our improved calculations.  This may be accomplished by recognizing that the analysis of \citet{dressing2013} used both the post-marginalized method of completeness correction and a SNR=7.1 detection threshold (a calculation in the style of the dotted line in Figure \ref{fig:rdistcompare}.  As the integral of our reconstructed radius function on the interval [0,1.4] $\Rearth$ is about 1.6$\times$ larger than the integral over the same range of the post-marginalized method using the SNR threshold, we conclude that there should be closer to 0.25 habitable-zone Earth-like planets per cool star, rather than the $\sim$0.15 estimated by that work. And if this same correction is made to the calculations of \citet{kopparapu2013b}, which use updated HZ calculations but the same occurrence formalism as \citet{dressing2013}, than this number would become closer to $\sim$0.8 planet per star.  It is likely that habitable-zone, Earth-sized planets abound throughout the Galaxy in numbers even larger than previously estimated.

Finally, we emphasize that this calculation is based on a target sample of only about 3900 cool stars and a KOI search only through Q12 data.  Future pipeline searches and continued observations of cool stars by the K2 mission, as well as future surveys such as TESS and PLATO, will increase this sample size, allowing for strengthened conclusions from the small-planet radius distribution and giving a greater handle on the formation processes of planetary systems around the most numerous stars in the Galaxy.  In addition, careful application of these same principles to the entire \Kepler\ dataset, as permitted by accurate knowledge of stellar parameters, will continue to uncover important clues to the formation and evolution of all types of planetary systems.

\acknowledgments{The authors acknowledge John Johnson and the Caltech Exolab for nurturing the intellectual environment in which this work took shape, in particular, conversations with Phil Muirhead, Leslie Rogers, Avi Shporer, Ben Montet, and Jean-Michel Desert.  TDM thanks Ed Turner and David Hogg for supportive conversations regarding this analysis, and Frank Porter for very helpful tips regarding density estimation, and for his willingness to share the draft version of his book \citep{narsky2013}, by which the bootstrap techniques used in this work were inspired.  We also thank Peter Goldreich for an enlightening conversation about rocky planet formation, Dave Charbonneau and Courtney Dressing for helpful comments on an early draft.  TDM acknowledges support from the \Kepler\ Participating Scientist Program, NASA NNX11AG85G.}

\bibliographystyle{apj}
\bibliography{allrefs}

\end{document}